\documentstyle[12pt]{article}

\newcommand{\bmat}{\left(\begin{array}}
\newcommand{\emat}{\end{array}\right)}

\def\yzero{\smash{\hbox{$y\kern-4pt\raise1pt\hbox{${}^\circ$}$}}}

\def\beq{\begin{equation}}
\def\eeq{\end{equation}}
\def\beqa{\begin{eqnarray}}
\def\eeqa{\end{eqnarray}}

\def\-{\hphantom{-}}

\def\s2{\frac{1}{2}}

\def\beq{\begin{equation}}
\def\eeq{\end{equation}}
\def\beqa{\begin{eqnarray}}
\def\eeqa{\end{eqnarray}}

\def\IF{\relax{\rm I\kern-.18em F}}
\def\II{\relax{\rm I\kern-.18em I}}
\def\IP{\relax{\rm I\kern-.18em P}}

\def\cp{{\cal P}}
\def\IC{\bf C}
\def\IZ{\bf Z}
\def\IR{\bf R}
\def\IS{\bf S}
\def\z2z2{$\IC^3/(\IZ_2\times\IZ_2)$}

\def\Dsl{\,\raise.15ex\hbox{/}\mkern-13.5mu D} 

\def\IT{\bf T}

 \def\cp#1{\relax\ifmmode {\IP\kern-2pt{}_{#1}}\else $\IP\kern-2pt{}_{#1}$\=fi}

\begin{document}

\makeatletter \@addtoreset{equation}{section} \makeatother
\renewcommand{\theequation}{\thesection.\arabic{equation}}
\pagestyle{empty}
\vspace*{.5in}
\rightline{CERN-TH/2001-233}
\rightline{\tt hep-th/0108196}
\vspace{2cm}

\begin{center}
\LARGE{\bf Wrapped fluxbranes \\[10mm]}
\medskip
\large{Angel M. Uranga \\[2mm]}
{Theory Division, CERN, CH-1211 Geneva 23, Switzerland.
\\[3mm]}

\smallskip

\small{\bf Abstract} \\[3mm]
\end{center}
We consider the construction of fluxbranes in certain curved geometries, 
generalizing the familiar construction of the Melvin fluxtube as a 
quotient of flat space. The resulting configurations correspond to 
fluxbranes wrapped on cycles in curved spaces. The non-trivial transverse 
geometry leads in some instances to solutions with asymptotically 
constant dilaton profiles. We describe explicitly several supersymmetric 
solutions of this kind. The solutions inherit some properties from their 
flat space cousins, like flux periodicity. Interestingly type IIA/0A 
fluxbrane duality holds near the core of these fluxbranes, but does not 
persist in the asymptotic region, precisely where it would contradict 
perturbative inequivalence of IIA/0A theories.

\begin{center}
\begin{minipage}[h]{14.5cm}
{\small

}

\end{minipage}
\end{center}
\newpage
\setcounter{page}{1} \pagestyle{plain}
\renewcommand{\thefootnote}{\arabic{footnote}}
\setcounter{footnote}{0}

\section{Introduction}
\label{intro}

The study of solitonic localized lumps of energy in string theory has 
turned out to provide far-reaching insights into its nature.
This has been the case for BPS D$p$-branes (or $p$-branes in 
general), and more recently for non-BPS D-branes. Recently, a different 
kind of objects, the fluxbranes, has started to receive attention
\cite{new1,new2,jorge,gs,gr,chen,cosgut,saffin1,fluxbranes,dielectric,roberto,brecher,saffin2,suyama,motl}. 
A flux $p$-brane (F$p$-brane) is a higher dimensional generalization of a 
flux-tube, namely a $(p+1)$-dimensional object with non-zero flux for a
degree $(9-p)$ field-strength form $F_{9-p}$ (or wedge product of field 
strength forms) in the transverse $(9-p)$-dimensional space. Hence the 
core of the fluxbrane carries a $F_{9-p}$ flux piercing the transverse 
space.

Remarkably, a non-trivial version of these objects, generalizing the 
Melvin fluxtube \cite{melvin}, can be constructed by 
imposing non-trivial identifications of points in (one dimension 
higher) flat space and performing a `skew' Kaluza-Klein reduction 
\cite{old2} \footnote{See \cite{fluxbranes,dielectric,saffin1,saffin2} 
for discussion on the direct constructions of general fluxbrane solutions in 
supergravity.}. The simple lift to the higher dimensional theory of this 
fluxbrane allows to study its properties in detail. For instance, its 
instability and decay modes \cite{old2,new1,new2}, duality between 
fluxbrane solutions in type IIA and 0A theories \cite{cosgut}, and duality 
with other string configuration \cite{fluxbranes}. It has also allowed to 
use such fluxbranes as sources to trigger Myers' dielectric effect on 
D-branes \cite{myers} and build new hopefully stable configurations 
\cite{dielectric,roberto,brecher}.

Unfortunately, the Melvin type solutions lead to blowing up dilaton values 
at large distance, which prevent for a more detailed understanding and 
interpretation of the solution from the lower dimensional perspective. 
Also, some directly constructed fluxbranes have non-normalizable transverse 
flux. In a sense these results stem from the fact that these fluxbranes live 
in flat space (which becomes curved after introduction of the flux). In this 
setup self-gravitation of the flux, although leads to the formation of a 
flux-brane, does not prevent it from spreading too much. One would expect 
that considering fluxbranes in a background gravitational field (to 
which self-gravity subsequently adds) would improve the above features.

The purpose of this paper is to discuss the construction of fluxbranes in 
certain curved spaces, by a simple generalization of the construction of 
the Melvin fluxbranes from flat space
\footnote{A different possibility, not explored in this paper, is 
to consider `composite' fluxbranes in flat space. Their core contains 
fluxes for different gauge fields, and may lead to asymptotically 
constant scalars, for suitably tuned fluxes \cite{jorge,roberto2}. We 
thank R.~Emparan for pointing out this possibility.}. In some situations, 
the flux is `confined' by the background transverse geometry, so that 
asymptotically in the transverse space the corresponding solutions have 
constant dilaton, and vanishing flux. We find that in this situation 
certain properties of the flat space solution, like flux periodicity and 
certain dualities, remain, while for instance IIA/0A fluxbrane duality is 
substantially modified.

These solutions also illustrate that fluxbranes are dynamical objects, 
able to wrap non-trivial cycles in curved spaces. Consequently they should 
contain world-volume zero modes associated to deformations of the cycles, 
and a rich dynamics which remains unexplored. 

The paper is organized as follows. In Section 2 we present a generalized 
version of the construction of e.g. type IIA fluxbranes in a spacetime 
${\bf X}_{10}$ by modding out ${\bf X}_{11}=\IR\times {\bf X}_{10}$ by a 
simultaneous shift in the additional dimension and an arbitrary $U(1)$ 
isometry in ${\bf X}_{10}$. In Section 3 we review the case of rotational 
isometries in flat space ${\bf X}_{10}$, which leads to the Melvin 
fluxbrane and some simple generalizations. We point out an unnoticed 
connection with fluxbranes at orbifold singularities. In Sections 4,5 we 
use our generalized construction to build fluxbranes wrapped on cycles in 
curved spaces. We present explicit examples of IIA F5-branes in Taub-NUT 
space, and on a 3-cycle in the $G_2$ holonomy manifold in \cite{bggg}. 
These solutions have asymptotically constant dilaton and vanishing flux. 
We also show that wrapping does not guarantee the latter features, and 
give examples of wrapped fluxbranes with blowing up dilaton. In Section 6 
we briefly discuss examples where the isometry in ${\bf X}_{10}$ has no 
fixed points, and construct a fluxbrane in an AdS compactification. In 
Section 7 we discuss IIA/0A fluxbrane duality in models with constant 
asymptotic dilaton. Finally, Section 8 contains our final remarks.

\section{General Strategy}

In this section we provide a generalization of the familiar procedure 
\cite{old2,new1,new2} to obtain fluxbranes in a $d$-dimensional 
theory as quotients of simpler configurations in $d+1$ dimensions. For 
concreteness we phrase the discussion in terms of IIA/M theory language, 
even though it is clear how to use the results to build e.g. NS-NS 
fluxbranes.

We consider M-theory on a space ${\bf X}_{11}=\IR\times {\bf X}_{10}$, 
with ${\bf X}_{10}$ admitting a $U(1)$ isometry parametrized by $\tau$. 
The 
general form of the metric is
\beqa
ds_{11}^{\,2} \, =\, dx_{10}^{\,2} + f(x)\, (d\tau + f_\mu(x)\,dx^\mu)^2 + 
g_{\mu\nu}\, dx^\mu\, dx^\nu
\label{metric}
\eeqa
where $x$ denotes the set of $x^{\mu}$'s, $\mu=0,\ldots, 8$, and $x$, 
$\tau$ parametrize ${\bf X}_{10}$, and $x_{10}$ parametrizes $\IR$. The 
isometry generated by the Killing vector $\partial_{\tau}$, and we 
normalize the period of $\tau$ to $2\pi$.

Consider modding by $x_{10}\to x_{10}+2\pi R_{10}$, $\tau \to \tau + 
2\pi B$. We are interested in performing a KK reduction along the Killing 
$R_{10}\partial_{10}+B\partial_\tau$. Defining the adapted coordinate 
${\tilde \tau}=\tau -\chi x_{10}$, with $\chi=B/R_{10}$ the metric becomes
\beqa
ds_{11}^{\,2} \, =\, dx_{10}^{\,2} + f(x)\, (d{\tilde\tau}+\chi\, 
dx_{10} \, =\, f_\mu\, dx^\mu)^2 + g_{\mu\nu}\, dx^\mu dx^\nu
\eeqa
In these coordinates, the KK reduction is simply along the Killing vector 
$\partial_{10}$. In order to obtain the ten-dimensional quantities after 
the reduction, we compare with the KK ansatz metric
\beqa 
ds_{11}^2 \; =\; e^{4\Phi/3} \, (dx_{10} + R_{10} A_\mu dx^\mu)^2 + 
e^{-2\Phi/3} ds_{10}^{\,2}
\label{kkansatz}
\eeqa
where $\Phi$ and $A$ are the ten-dimensional dilaton and RR 1-form 
fields, respectively. After some algebra, one obtains
\beqa
& e^{4\Phi/3}= \Lambda(x) \quad ; \quad 
R_{10} A = \Lambda^{-1} f(x)\, \chi\, (d{\tilde\tau}+f_\mu\, dx^\mu) & 
\nonumber\\
\\
& ds_{10}^{\,2} = \Lambda^{-1/2}\, f(x)\, (d{\tilde\tau} + f_\mu\, 
dx^\mu)^2 + \Lambda^{1/2}\, (\,g_{\mu\nu}\, dx^\mu\, dx^\nu + 
d\vec{x}^{\,2}) & \nonumber
\label{tendim}
\eeqa
with $\Lambda=1+ \chi^2 f(x)$.

The resulting ten-dimensional solution has non-zero RR 1-form field 
strength, related to the variations of the geometric features of the 
original $U(1)$ isometry orbits in ${\bf X}_{10}$. When the resulting 
field strength configuration forms a lump, the solution is interpreted as 
a fluxbrane, generalizing the familiar concept of fluxtubes. 

The prototypical case of these solutions, reviewed in next section, is 
given by starting with flat space ${\bf X}_{11}= \IR\times {\bf M}_{10}$, 
and using a rotational isometry in ${\bf M}_{10}$. The size of the $U(1)$ 
orbits increases steadily as one moves away from the fixed locus of the 
rotation, which implies the resulting solution is a fluxbrane spanning 
the (image after KK reduction of the) fixed plane. Unfortunately, the 
growing of the orbit size has no bound, and leads to a unbounded dilaton 
growth driving the final solution to strong coupling at large distances 
away from the fluxbrane.

New more general possibilities, leading to wrapped branes of different 
kinds, are discussed in subsequent sections.

\section{Flat space fluxbranes}
\label{flat}

In this section we apply the strategy above to review the construction of 
Melvin fluxbranes (and generalizations) in flat space, essentially 
following \cite{new2}.

\subsection{The IIA F7-brane}
\label{flatconst}

Let us review the construction of the Melvin fluxbrane solution in IIA 
string theory starting from flat space in M-theory \footnote{Our conventions 
differ slightly from other references.}. Consider M-theory in flat space, 
with metric
\beqa
ds_{11}^2 = dx_{10}^2 + d\rho^2 + \rho^2 d\varphi^2 + d\vec{x}^{\,2}
\eeqa
where $\rho, \varphi$ are polar coordinates in an $\IR^2$ and $\vec{x}\in 
{\bf M}^8$. Consider performing the identification
\beqa
x_{10} & \to & x_{10} + 2\pi R_{10} \nonumber \\
\varphi & \to & \varphi + 2\pi B
\label{identfseven}
\eeqa
Namely, a $2\pi$ shift identification along the Killing 
$R_{10}\partial_{10} + B \partial_{\varphi}$. This action breaks all the 
supersymmetries of the configuration. Comparing with (\ref{metric}) we have
\beqa
&\tau=\varphi \quad ; \quad f=\rho^2 \quad ; \quad f_\mu=0  
\quad ; \quad  g_{\mu\nu}\, dx^\mu\, dx^\nu = d\rho^2 + d\vec{x}^{\,2} &
\eeqa
Applying those results, we define the adapted coordinate ${\tilde 
\varphi}= \varphi - (B/R_{10}) x_{10}$,
and replacing into (\ref{tendim}) we obtain
\beqa
e^{4\Phi/3} = 1+\frac{B^2 \rho^2}{R_{10}^{\,2}} \quad\quad ; \quad\quad 
A_{\tilde \varphi} = \frac {B\rho^2}{R_{10}^{\,2}+B^2\rho^2} \nonumber \\
\\
ds_{10}^{\,2} = \Lambda^{1/2}\, d\rho^2 + \Lambda^{-1/2} \rho^2\, 
d{\tilde \varphi}^2 + \Lambda^{1/2}\, d\vec{x}^{\,2}
\eeqa
with $\Lambda=1+B^2 \rho^2/R_{10}^{\,2}$. The solution describes a 
8-dimensional Poincare invariant configuration of non-vanishing RR 1-form 
field strength flux turned on in the $(\rho,{\tilde\varphi})$ plane. Near 
the origin the metric is smooth and has small curvature, and the RR field 
strength is approximately constant. This configuration is the type IIA 
F7-brane, originally discussed in \cite{gs}. At large $\rho$ the 
fluxbrane induces strong coupling, thus affecting the validity of the 
description. This property makes the interpretation of the solution 
difficult in certain regimes. 
As discussed in the introduction, the root of this problem lies in the use 
of isometries with growing orbit length in the base space. 

The weakly coupled region is at $\rho\ll R_{10}/B$, while the ten-dimensional 
approximation hols for $\rho\gg R_{10}$. Hence, for both conditions to 
hold one need $B\ll 1$, weak field strength. Below we discuss certain 
duality properties, which typicaly involve strong/weak field strength 
relations, and therefore imply either strong coupling or breakdown of the 
ten-dimensional approximation.

\subsection{Lower dimensional F-branes}

For completeness, let us discuss the construction of other fluxbranes with 
non-vanishing flux for the $2k$-form $F\wedge \ldots \wedge F$ in their 
$(10-2k)$-dimensional core, denoted F$(9-2k)$-branes. Let us write the 
11-dimensional flat metric as
\beqa
ds_{11}^2 = dx_{10}^{\,2} + \sum_{i=1}^k \,( \,d\rho_i^2 + \rho_i^2\, 
d\varphi_i^2) + d\vec{x}^{\,2}
\eeqa
where $\rho, \varphi$ are polar coordinates in the $i^{th}$ $\IR^2$ and 
$\vec{x}\in {\bf M}^{10-2k}$. Let us perform the identification
\beqa
x_{10} & \to & x_{10} + 2\pi R_{10} \nonumber \\
\varphi_i & \to & \varphi_i + 2\pi B_i
\label{identgen}
\eeqa
Equivalently, defining complex coordinates $z_j=\rho_j e^{i \varphi_j}$, 
the action on $\IR^{2k}$ is $z_j\to e^{iB_j}z_j$. For generic choices of 
$B_i$ this action breaks all the supersymmetries of the configuration. 
However, for actions on $\IR^{2k}$ lying in a subgroup of $SU(k)$ some 
supersymmetry is preserved.

In this case it is simpler to repeat the calculation instead of using the 
result in Section 2. Defining the adapted coordinates 
\beqa
{\tilde \varphi_i}= \varphi_i - \frac{B_i}{R_{10}} x_{10}
\eeqa
the metric becomes
\beqa
ds_{11}^2 = dx_{10}^2 + \sum_i\, [\, d\rho_i^2 + \rho_i^2\, (d{\tilde 
\varphi}_i 
+\frac{B_i}{R_{10}} dx_{10})^2 \, ] + d\vec{x}^{\,2}
\eeqa
Performing the KK reduction along $\partial_{10}$, and comparing with
(\ref{kkansatz}). We obtain
\beqa
& e^{4\varphi/3} = \Lambda \quad \quad ; \quad \quad 
A_{{\tilde \varphi}_i} = B_i\,\rho_i^2\, / \,
(R_{10}^2\, + \sum_{j=1}^k\, B_j^2\, \rho_j^2) & \\
& ds_{10}^{\, 2} = \Lambda^{1/2} \sum_{i=1}^k (\, d\rho_i^2 + \rho_i^2\, 
d{\tilde \varphi}_i^2 \,) +\Lambda^{1/2}\, d\vec{x}^{\,2} - \Lambda^{-1/2} 
[\, \sum_{i=1}^k (B_i\, \rho_i^2\, /\, R_{10}) \, d{\tilde \varphi}_i\,]^2 
&
\nonumber
\eeqa
with $\Lambda=1+\sum_{i=1}^k B_i^2 \rho_i^2/R_{10}^{\,2}$. The 
solution describes a $(10-2k)$ Poincare invariant object with nonzero 
$F^k$-flux in the transverse $\IR^{2k}$, a F$(9-2k)$-brane. In analogy 
with the previous situation, the dilaton blows up away from the core of 
the fluxbrane.

\subsection{Dualities}

We conclude this section by reviewing some of the remarkable properties 
of these solutions. Although they are very surprising from the 
10-dimensional viewpoint, they are readily obtained from the above 
11-dimensional construction, by the simple strategy of choosing 
KK reductions with respect to different Killing isometries.

A first such property is flux periodicity. Namely, the  fluxbrane 
configurations for parameters $B$ and $B+2$ are clearly equivalent, since 
they have the same 11-dimensional lift (since a rotation by $4\pi$ is 
trivial). Concretely, starting with flat space and the identification
(\ref{identfseven}) reduction along $R_{10}\partial_{10}+B\partial_\varphi$ 
leads to the IIA F7-brane with parameter $B$, while reduction along
$R_{10}\partial_{10}+(B+2)\partial_\varphi$ leads to the F7-brane with 
parameter $B+2$. The equivalence is however highly non-trivial from the 
viewpoint of the 10-dimensional variables.

A second duality property is derived analogously. Start again with the 
quotient of flat space by the identification (\ref{identfseven}). Reducing 
along the Killing $R_{10}\partial_{10} + B \partial_{\varphi}$ we obtain the 
IIA F7-brane. On the other hand, reducing along the killing 
$R_{10}\partial_{10} + (B+1) \partial_{\varphi}$, fermions are 
antiperiodic along the corresponding $\IS^1$ orbit (due to the additional 
$2\pi$ rotation in the $(\rho,\varphi)$ plane). As pointed out in 
\cite{cosgut}, given the IIA/0A/M-theory duality proposal in \cite{bg}, 
the resulting F7-brane is embedded in type 0A theory. Hence one obtains a 
duality between the type IIA and 0A  F7-branes, for different  values of 
the flux.

Considering the same starting point, the IIA F7-brane can be related to yet 
another familiar configuration in cases where $B=1/N$. Namely, reducing the 
11-dimensional configuration along the Killing $\partial_{10}$, the 
resulting configuration is simply an orbifold flat 10-dimensional space, 
with orbifold action generated by
\beqa
z\to e^{2\pi i \, B} \,z
\eeqa
where $z=\rho\, e^{i\varphi}$. This action is moreover embedded into the 
RR 1-form gauge degree of freedom, as in \cite{schwarzsen}. This is the 
F7/cone duality \cite{fluxbranes}.

Analogously the $F(9-2k)$-branes can be shown to be dual, for suitable 
$B_i$, to $\IC^{k}/\IZ_N$ orbifold singularities. In fact, an interesting 
connection has been proposed between the generic instabilities of 
non-supersymmetric fluxbranes \cite{new1,new2} and the closed string 
twisted tachyons in non-supersymmetric orbifold singularities 
\cite{panic} (see also \cite{jorge}).

\subsection{Fluxbranes at singularities}

These dualities are straightforwardly extended to lower-dimensional  
fluxbranes. In the following we consider a particular duality not 
explicitly discussed in the literature, and which leads to interesting 
systems of fluxbranes at orbifold singularities.

In particular, consider flat 11-dimensional space modded out by the 
identification (\ref{identgen}). Reducing along the Killing $R_{10}\,
\partial_{10} + \sum_{i=1}^k B_i\, \partial_{\varphi_i}$ leads to a 
F$(9-2k)$-brane in flat space. Consider the case in which a subset of the 
$B_i$ are rational numbers $a_i/N$. We may index such entries 
by $i=1,\ldots, k_1$, and the remaining $i'=1,\ldots,k_2$, with 
$k_1+k_2=k$. Reducing along the Killing $R_{10}\,\partial_{10}+
\sum_{i'=1}^{k_2} B_{i'}\, \partial_{\varphi_{i'}}$, we obtain a 
F$(9-2k_2)$-brane sitting transverse to a $k_1$-fold singularity, 
generated by
\beqa
(z_1,\ldots,z_{k_1})\to (e^{2\pi \frac{a_i}{N}} z_1,\ldots, 
e^{2\pi \frac{a_{k_1}}{N}} z_{k_1})
\eeqa
with the twist embedded in the RR 1-form gauge degree of freedom i.e. a 
flat connection RR field turned on, as in \cite{schwarzsen}. The $F^{k_2}$ 
flux is turned on in the $\IR^{2k_2}$ directions transverse to the 
fluxbrane and the orbifold $\IC^{k_1}/\IZ_N$. Interestingly, for certain 
choices of the $B$'s (which determine the fluxbrane species and the 
orbifold action) the system of fluxbranes at the singularity is 
supersymmetric, even if neither of them is supersymmetric independently. 
From this viewpoint, the introduction of fluxbranes at non-supersymmetric 
orbifold singularities provides a stabilization 
mechanism for the latter (and vice-versa). It would be interesting to 
understand this from an analysis similar to \cite{panic}.

\medskip

Even though the above dualities are suggestive, one cannot avoid the 
feeling that their usefulness (or even validity) is questioned by the 
blowing up behaviour in the large distance regime. It would 
be desirable to have a similarly simple construction of fluxbranes in a 
more controlled setup, with asymptotically constant dilaton and small 
curvature. In the following sections we provide such a construction, 
which leads to fluxbranes wrapped on cycles in a curved space. 
Intuitively, the non-trivial transverse geometry cuts off the flux at 
large distances, so that asymptotically the backreaction of the flux on 
the geometry and the dilaton gradient fall off. 

\section{Fluxbranes in Taub-NUT space}
\label{taubnut}

In this section we discuss the construction of a fluxbrane with constant 
asymptotic dilaton. It has the interpretation of a fluxbrane transverse to 
a Taub-NUT space. 

Our starting point is the $N$-center Taub-NUT space \cite{tn}, defined by 
the metric
\beqa
ds^2 & = &  U^{-1}\, (d\tau + \vec{\omega} \cdot d\vec{r}\,)^{\,2} + 
U\,  d\vec{r}^{\,2} \nonumber \\
U & = & \frac{1}{R^2} + \sum_{i=1}^N \, \frac{1}{|\vec{r}-\vec{r}_i|} 
\quad ; \quad \vec{\nabla}\times  \vec{\omega} =\vec{\nabla} U
\label{taub}
\eeqa
The coordinate $\tau$ has periodicity $2\pi$. This space is an $\IS^1$ 
fibration over $\IR^3$, with fiber of constant asymptotic radius and 
degenerating over the points $\vec{r}_i$, the centers. The metric is 
smooth as long as the latter are generic, and develops $A_{k-1}$ orbifold 
singularities when $k$ centers coincide.

The metric is hyperkahler, therefore has an $SU(2)$ isometry. This is most 
manifest for coinciding centers, where it amounts to rotations on the 
base $\IR^3$. The metric also has a $U(1)$ isometry along the $\IS^1$ 
fiber, corresponding to shifts in $\tau$. The orbits of this isometry 
have asymptotically constant length $2\pi R$, and degenerate over the 
centers. It is this isometry that we will exploit to build our fluxbrane.

Let us consider compactifying M-theory in Taub-NUT space, a configuration 
which preserves sixteen supersymmetries. Denoting the additional coordinates 
$x_{10}$ and $\vec{x}\in \IR^6$, the metric is
\beqa
ds_{11}^{\,2} \, = \, U^{-1}\, (d\tau + \vec{\omega} \cdot 
d\vec{r}\,)^{\,2} + U \, d\vec{r}^{\,2} + dx_{10}^{\,2} + d\vec{x}^{\,2}
\eeqa
Let us identify points related by the action $x_{10} \to x_{10} + 2\pi 
R_{10}$, $\tau \to \tau + 2\pi R\, B$, namely points related by a $2\pi$ 
shift along the Killing direction $R_{10}\partial+ RB \partial_\tau$. 
This identification preserves all sixteen supersymmetries in the system. 

Comparing with (\ref{metric}), we have
\beqa
&f=U^{-1} \quad ; \quad f_{\vec{r}}=\vec{\omega}  \quad ; \quad 
f_{\vec{x}}=0 \quad ; \quad
 g_{\mu\nu} dx^\mu dx^\nu = U \, d\vec{r}^{\,2} + d\vec{x}^{\,2}&
\eeqa
We introduce the adapted coordinate
\beqa
{\tilde \tau} = \tau - \chi\, x_{10}, \quad {\rm with} \;\;
\chi=\frac{RB}{R_{10}}
\eeqa
and working directly, or using (\ref{tendim}), the KK reduction gives
%
\beqa
&e^{4\Phi/3}  =  \Lambda \quad \quad ;\quad \quad
R_{10}\, A  =  \chi\, \Lambda^{-1} U^{-1} (\,d{\tilde \tau} 
+ \vec{\omega} \cdot d\vec{r}\,) & \\
& ds_{10}^{\,2}  =  \Lambda^{-1/2}\, U^{-1}\,  (d{\tilde \tau} + 
\vec{\omega} \cdot d\vec{r}\,)^{\,2}  +  \Lambda^{1/2}\, U\, 
d\vec{r}^{\,\,2} + \Lambda^{1/2}\, d\vec{x}^{\,2} 
& \nonumber
\eeqa
with $\Lambda=1+\chi^2\, U^{-1}$.

The geometry is similar to that of Taub-NUT space, with a `squashing' 
which is important near the centers, and dies off for large $r$. The 
systems is however supersymmetric due to the presence of a non-trivial RR 
1-form flux, again localized near the centers $\vec{r}_i$, and asymptoting 
to a flat connection. Finally the dilaton is well behaved everywhere, and 
asymptotes to a constant value. The metric asymptotically tends to a 
standard Taub-NUT compactification, with redefined radius due to the 
new proper length along the orbit of ${\tilde\tau}$. The $U(1)$ isometry 
in the neighbourhood of the Taub-NUT center, is locally of the form
\beqa
(z_1,z_2) \to (e^{i\alpha}z_1,e^{-i\alpha} z_2)
\label{isom}
\eeqa
for suitable complex coordinates $z_1,z_2$ \footnote{Near a center the 
Taub-NUT metric is $ds^2 = \frac 1r\, d\vec{r}^{\,2} + r\,(d\tau + 
\vec{\omega}\cdot d\vec{r})^{\,2}$. Changing variables as $q=ae^{i\tau}$, 
where $a$ is a purely imaginary quaternion, the metric becomes manifestly 
flat $ds^2=dq d{\bar q}$, and the $U(1)$ isometry becomes (\ref{isom}) in 
terms of $q=z_1 + j z_2$, see \cite{gibrych} for details.}. Hence, we 
obtain non-zero $F^2$ flux localized at those points. At large $r$, the 
isometry becomes translational, and the KK reduction simply reproduces a 
flat connection, a RR 1-form Wilson line. Since the flux dies off 
asymptotically, one recovers a Taub-NUT space and a constant dilaton 
in the large $r$ region.

The solution therefore corresponds to an F5-brane transverse to a Taub-NUT 
geometry. The final configuration preserves sixteen supersymmetries, the 
same amount as a Taub-NUT geometry with no flux. Hence in this case the 
fluxbrane does not break any further supersymmetries.

An advantage of the good asymptotics of this solution is that the 
fluxbrane can be understood as an excitation of type IIA theory on 
Taub-NUT space. As such, one may hope to learn about the existence of zero 
modes of the excitation, localized on its core, by looking at 
symmetries preserved by the background and broken by the excited state, 
namely Goldstone bosons or goldstinos. Unfortunately, the fluxbrane does 
not break any further symmetries, hence no such modes arise. This 
does not contradict the claim that fluxbranes are dynamical; even for 
D-branes, there exist cases (like branes trapped at singularities) where 
no Goldstone zero modes exist. Hence our solution describe a trapped 
fluxbrane in the same sense.

We refrain from discussing duality properties of this solution and 
variants thereof, which are easily derived from the 11-dimensional 
perspective by applying the philosophy in \cite{fluxbranes}. Instead, we 
turn to the construction of other interesting examples.

\section{More general wrapped fluxbranes}
\label{moregen}

\subsection{Wrapped fluxbranes with asymptotically constant dilaton}
\label{constant}

In this section we would like to further explore the above ideas to 
construct examples of fluxbranes with constant asymptotic dilaton, 
and wrapped on diverse cycles in non-trivial geometries. In order to get 
constant asymptotic dilaton, it is crucial that the base spaces have 
$U(1)$ isometries with orbits of asymptotic constant length. 
Metrics for such spaces are not easy to construct in dimensions higher 
than four, hence in this section we use an indirect route to consider 
such geometries. 

The key idea is that such geometries are predicted to exist by string 
duality. Specifically, the lift to M-theory of configurations involving 
D6-branes corresponds to non-trivial geometries with a $U(1)$ isometry of 
constant asymptotic length (related to the asymptotic IIA dilaton). For 
supersymmetry preserving D6-brane configurations, a nice list of the 
special holonomies arising in M-theory is given in \cite{jaume}. We will 
use this duality relation to generate spaces on which we subsequently 
wrap fluxbranes.

\medskip

{\bf General examples}

Let us construct e.g. a F5-fluxbrane wrapped on a holomorphic 2-cycle on a 
Calabi-Yau threefold. We will be interested in discussing CY spaces with a 
$U(1)$ isometry of constant asymptotic radius orbit. A simple way to 
generate such spaces is to use string duality, as follows. Consider a 
D6-brane wrapped on a (possibly non-compact) holomorphic 2-cycle 
${\bf \Sigma}_2$ within a (possibly non-compact) four-dimensional 
hyperkahler space ${\bf X}_4$. Lifting the configuration results in a 
Calabi-Yau threefold ${\bf X}_6$, topologically a $\IC^*$ fibration over 
${\bf X}_4$, with fibers degenerating (to two intersecting complex 
planes) over ${\bf \Sigma}_2$. The lift also provides us with some metric 
information, namely that the $\IS^1$ fiber within the $\IC^*$ is the 
orbit of a $U(1)$ isometry, has constant asymptotic radius (the radius of 
the M-theory circle), and degenerates over ${\bf \Sigma}_2$. This space 
is a threefold generalization of the situation encountered in Section 4. 
In fact, the full geometry is just a fibration of the Taub-NUT space over 
${\bf \Sigma}_2$.

Let $\tau$ parametrize the $\IS^1$ fiber, with period $2\pi$, and 
$x_{10}$ 
parametrize one of the additional flat non-compact dimensions. Consider 
modding out by the symmetry $x_{10} \to x_{10}+2\pi R_{10}$, $\tau \to\tau 
+ 2\pi B$, and perform a KK reduction along the Killing vector $R_{10}\,
\partial_{10} +B\,\partial_\tau$. Since at every point in ${\bf \Sigma}_2$ 
the 
geometry is locally like in the Taub-NUT example, one obtains an F5-brane
spanning four flat non-compact dimensions and wrapped on ${\bf \Sigma}_2$ 
within (a squashed version of) ${\bf X}_6$. The construction preserves 8 
supercharges, the same amount preserved by ${\bf X}_6$ alone. So the 
fluxbrane does not break further supersymmetries. 

Notice that if ${\bf \Sigma}_2$ is not rigid the configuration contains 
zero modes associated to deformations of the cycle. These however exist 
already in the geometry of ${\bf X}_6$, and hence should not be assigned 
to the fluxbrane. Again, the fluxbrane is trapped at the core of the 
Taub-NUT fibers, and is simply dragged by the deformations of the 
geometry.

\medskip

The construction is obvious to generalize to similar fluxbranes on diverse 
cycles. Sticking to supersymmetric cycles in special holonomy manifolds, 
many examples can be generated using string duality. Without repeating 
details, one may build: 1) Starting with a D6-brane on a Slag 3-cycle 
${\bf \Sigma}_3$ within a Calabi-Yau threefold ${\bf X}_6$, we get a 
F5-brane on a 3-cycle  in a (squashed version of a) $G_2$ manifold; 2) 
Starting with a D6-brane on a holomorphic 4-cycle ${\bf \Sigma}_4$ in a 
Calabi-Yau threefold, an F5-brane on a holomorphic 4-cycle in a Calabi-Yau 
fourfold; 3) Starting with a D6-brane on a coassociative 4-cycle in a 
$G_2$ manifold, we get an $F^2$ fluxbrane in a 4-cycle in a  $Spin(7)$ 
manifold; {\rm etcetera}.

It should be clear that these examples generate F5-branes due to the 
special nature $U(1)$ fixed loci in supersymmetry preserving manifolds, 
which are most familiar. Clearly, use of other isometries can lead to 
e.g. wrapped F7-branes using analogous techniques (see Section 5.2 and 6 
for some discussion).

\medskip

{\bf F5-brane in a 3-cycle in a $G_2$ manifold}

In the following we consider an example of the above construction. 
Recently a $G_2$ holonomy metric manifold with a $U(1)$ isometry with 
orbits of asymptotically constant radius has been constructed in 
\cite{bggg}. In the above language, it corresponds to the M-theory lift of 
a D6-brane wrapped on the 3-sphere in a deformed conifold. However, once 
the manifold metric is known, it can be used to build a fluxbrane without 
further appeal to how it was obtained, as we do below.

The metric for the compactification of M-theory in such space is 
\beqa
& ds_{11}^{\,2} \, = \, \sum_{a=1}^7 e^a \otimes e^a + d\vec{x}^{\,2}& 
\\
\nonumber \\
& e^1 = A(r)\; (\sigma_1-\Sigma_1) \quad ; \quad 
e^2 = A(r)\; (\sigma_2-\Sigma_2) & \nonumber \\
&e^4 = B(r)\; (\sigma_1+\Sigma_1) \quad ; \quad 
e^5 = B(r)\; (\sigma_2+\Sigma_2) & \nonumber \\
& e^3 = D(r)\; (\sigma_3-\Sigma_3) \quad ; \quad 
e^6 = r_0 C(r)\; (\sigma_3+\Sigma_3) & \nonumber \\
&e^7 =dr/C(r) &
\label{gtwo}
\eeqa
where $\vec{x}\in \IR^3$, and $\sigma_i$, $\Sigma_i$ are $SU(2)$ 
left-invariant 1-forms in $\IS^3$
\beqa
\begin{array}{ccc}
\sigma_1= \cos \psi\, d\theta + \sin \psi\, \sin \theta\, d\phi & \quad ; 
\quad 
& \Sigma_1= \cos {\hat\psi}\, d{\hat\theta} + \sin {\hat\psi}\, \sin 
{\hat\theta}\, d{\hat\phi} \\
\sigma_2= -\sin \psi\, d\theta + \cos \psi\, \sin \theta\, d\phi & \quad ; 
\quad
& \Sigma_2= -\sin {\hat\psi}\, d{\hat\theta} + \cos {\hat\psi}\, \sin 
{\hat\theta}\, d{\hat\phi} \\
\sigma_3= d\psi + \cos \theta\, d \phi & \quad ; \quad &
\Sigma_3= d{\hat\psi} + \cos {\hat\theta}\, d{\hat\phi} 
\label{sigmas}
\end{array} 
\eeqa
and $A$, $B$, $C$, $D$ are radial functions, explicitly given in eq.(3.6) 
of \cite{bggg}. The metric has isometry $SU(2)\times SU(2)\times U(1)$, 
with the $U(1)$ corresponding to a simultaneous shift in $\psi$, 
${\hat\psi}$ 
(which also induces a simultaneous rotation of $(\sigma_1,\sigma_2)$ and 
$(\Sigma_1,\Sigma_2)\,$). The orbit length is controlled by $C(r)$, and
asymptotes to constant $4\pi r_0$ at large $r$, and shrinks to zero at 
$r=9r_0/2$, which corresponds to an associative 3-cycle with 
$\IS^3$ topology \cite{bggg}. In fact, the full space can be thought of 
as a fibration of Taub-NUT over such cycle.

Defining $\tau=\psi+{\hat\psi}$, we may cast (\ref{gtwo}) in the form
\beqa
ds_{11}^{\,2} = d\vec{x}^{\,2} + \sum_{a\neq 6}  e^a \otimes e^a + r_0^2\, 
C(r)^2\, (d\tau + \cos\theta\, d\phi + \cos {\hat\theta}\, 
d{\hat\phi})^{\,2}
\eeqa
analogous to (\ref{metric}). Notice that the $\sum_a e^a e^a$ term 
is actually $\psi$-independent. Let us mod out this configuration by the 
isometry
\beqa
& x_{10}\to x_{10}+2\pi R_{10} \quad\quad ; \quad\quad  
\tau = \tau + 4\pi B 
\eeqa
We define ${\tilde \tau}=\tau - \chi\, x_{10}$, with $\chi=2 B/R_{10}$ and 
apply (\ref{tendim}) with
\beqa
f=r_0^2\, C(r)^2 \quad ; \quad f_{\phi}=\cos \theta \quad ; \quad 
f_{\hat\phi}=\cos{\hat\theta} \quad ; \quad g_{\mu\nu}=\sum_{a\neq 6} 
e^a\otimes e^a + d\vec{x}^2
\eeqa
We get a ten-dimensional IIA configuration given by
\beqa
 e^{4\Phi/3} & = & \Lambda  \quad ; \quad \quad 
\Lambda= 1+\chi^2 \, r_0^2 \, C(r)^2 \nonumber \\
A & = & \Lambda^{-1} \, r_0^2\, C(r)^2 \, (\chi/R_{10})\,
(\,d{\tilde\tau} +\cos\theta \, d\phi + \cos{\hat\theta}\, d{\hat\phi}\, ) 
\\
ds_{10}^{\, 2} & = & \Lambda^{1/2}\, d\vec{x}^{\,2} +\Lambda^{1/2} 
\sum_{a=1}^6  e^a \otimes e^a + \Lambda^{-1/2} \, r_0^2\, C(r)^2 
(\,d{\tilde\tau} +\cos\theta \, d\phi + \cos{\hat\theta}\, 
d{\hat\phi}\,)^{\, 2} \nonumber
\eeqa
The solution, by the above arguments, corresponds to an F5-brane wrapped 
on a 3-sphere in (a squashed version of) the $G_2$ manifold in 
\cite{bggg}.

\medskip

An a priori simpler realization of the same idea would be provided by 
starting e.g. with a D6-brane wrapped on for instance the 2-sphere in the 
Eguchi-Hanson space. Its lift would correspond to a Calabi-Yau three-fold 
${\bf X}_6$ given by a Taub-NUT fibration over such 2-cycle. Using this 
space to build a fluxbrane as described above would result in a 
supersymmetric  F5-brane wrapped on a 2-cycle in (an squashed version of) 
${\bf X}_6$. Even though this configuration is in principle simpler than 
the $G_2$ example above, the metric for ${\bf X}_6$ with asymptotically 
constant $\IS^1$ radius is not available in the literature. 

\medskip

We conclude by pointing out that these fluxbranes present flux periodicity 
by the same argument as in flat space. This supports the expectation of 
this being a generic feature of fluxbranes \cite{fluxbranes}. As in the 
flat case, our models with F5-branes do not show any connection between 
IIA and 0A solutions. Whether wrapped F7-branes do provide such connection 
or not is addressed in Section 7.

\subsection{Wrapped fluxbranes with blowing dilaton}

The remarkable property that the above fluxbranes have asymptotically 
constant dilaton is not generic within the class of wrapped fluxbranes. In 
other words, wrapping a fluxbrane does not guarantee that the dilaton is 
bounded at large distances. This is obvious e.g. in the above examples, 
by taking limits in which the asymptotic $U(1)$ circle decompactifies. 
For completeness we present some examples of wrapped fluxbranes with 
blowing dilaton.

\medskip

{\bf More fluxbranes at singularities}

Considering for instance F5-branes transverse to the $k$-center Taub-NUT 
case in Section 4. The original Taub-NUT space (\ref{taub}), in the limit 
$R\to \infty$ becomes an ALE space. For coincident centers, it is simply a 
$\IC^2/\IZ_k$, with the orbifold action generated by $(z_1,z_2) \to
(e^{2\pi i/k} z_1, e^{-2\pi i/k} z_2)$. Hence this construction yields a 
different configuration of fluxbranes at singularities, namely F5-branes 
transverse to $\IC^2/\IZ_N$, which is simply an orbifold of the flat space 
F5-brane solutions. Clearly, these solutions have large dilaton at large 
distances.

As an interesting aside, let us point out that using ideas similar to 
\cite{fluxbranes}, one readily derives a duality between fluxbranes in 
flat space and fluxbranes at singularities. In particular, consider  
M-theory in flat 11-dimensional space, with identification of points 
related by
\beqa
x_{10} \to x_{10} + 2\pi R_{10} \quad ; \quad
\varphi_1 \to \varphi_1 +2\pi B \quad ; \quad 
\varphi_2 \to \varphi_2 -2\pi B 
\eeqa
Reducing to ten dimensions along the Killing $R_{10}\partial_{10} + B 
\partial_{\varphi_1} - B\partial_{\varphi_2}$ lead to a F5-brane in flat 
space. On the other  hand, reducing along $R_{10}\partial_{10} + 
(B+1/k)\partial_{\varphi_1}-(B+1/k) \partial_{\varphi_2}$, leads to 
F5-branes for a different value of the flux, and transverse to a 
$\IC^2/\IZ_k$ singularity. The latter are precisely the same branes at 
singularities encountered above.

\medskip

{\bf Intersecting F7-branes in Taub-NUT}

To show that wrapped F7-branes can be achieved in our setup, let us build 
e.g. a F7-brane wrapped on a 2-cycle in multi Taub-NUT space, by using 
the $SU(2)$ (rather than the $U(1)$) isometry of this space. Considering 
the case of coincident centers, any $SU(2)$ element amounts to an $SO(2)$ 
rotation along some symmetry axis, say $x_3$, passing though the center 
location. Let us mod out by a $2\pi R_{10}$ shift in $x_{10}$ and a 
simultaneously $2\pi B$  rotation around $x_3$. Such identification breaks 
all the supersymmetries in the configuration. The result after KK 
reduction yields a RR 1-form flux localized on  a 2-dimensional subspace 
of Taub-NUT, parametrized by $x_3$ and $\tau$. Hence it corresponds to an 
F7-brane wrapped on a 2-cycle in 
Taub-NUT space, which is holomorphic in a suitable complex structure. As 
a complex space, Taub-NUT with coinciding centers can be described by 
$xy=v^N$. Adapting the coordinates, the rotational isometry corresponds 
to $x \to e^{i\alpha} x, y\to e^{i\alpha} y, v\to e^{2i\alpha/N} v$ for 
real $\alpha$. The flux is localized on its fixed points, which 
corresponds to the holomorphic 
(reducible) cycle $xy=0$. Hence we obtain two F7-branes wrapped on 
intersecting holomorphic cycles. The wrapping a non-supersymmetric object 
on a supersymmetric cycle explains the breaking of all supersymmetries.

\medskip

{\bf F5-branes in 2-cycle in the conifold}

Our final example is an F5-brane wrapped on a (non-compact) 2-cycle in 
the resolved conifold. Consider the resolved conifold, which is the 
cotangent bundle space of the 3-sphere ${\bf T^*S}^3$. The explicit metric 
can be found in \cite{philip}, but for our purposes it is enough to know that 
it admits an $SU(2)^2$ isometry. As a complex manifold the space can be 
described by the equation $z_1^2+z_2^2+z_3^2+z^4=\epsilon$, where epsilon 
is the deformation parameter, and the coordinates  $z_i$ transform in the 
$(2,2)$ representation of the $SU(2)^2$ isometry. A generic $U(1)$ 
rotation within $SU(2)^2$ acts as
\beqa
x\to e^{i\alpha} x \quad ; \quad  y\to e^{-i\alpha} y \nonumber \\
z\to e^{i\beta} z \quad ; \quad  w\to e^{-i\beta} w
\label{freeisom}
\eeqa
in adapted complex coordinates in which the geometry is described by 
$xy-zw=\epsilon$. Here $\alpha$ and $\beta$ define the embedding of $U(1)$ 
in the $U(1)^2$ within $SU(2)^2$.

Since the initial $SO(4)$ is the isometry of the 3-sphere in the 
resolution, the orbits of these symmetries grow with the distance. 
To build a fluxbrane, consider e.g. modding by the a simultaneous shift in 
$x_{10}$ and a $U(1)$ rotation with $\beta=0$. After KK reduction, we 
obtain a supersymmetric F5-brane wrapped on $x=y=0$, namely the 
holomorphic 2-cycle $zw=\epsilon$, which has the topology of a cylinder 
(it intersects the base $\IS^3$ over a maximal $\IS^1$). 

Alternatively, we could have chosen an isometry with non-zero $\alpha$, 
$\beta$. This isometry is freely acting on the deformed conifold, a 
situation not encountered before. Examples of this kind are briefly 
discussed in Section~6. For the undeformed conifold, $xy-zw=0$, this 
action leaves the origin as the only fixed point. The resulting fluxbrane 
is a supersymmetric F3-brane transverse to the conifold singularity.

Let us conclude by pointing out that these examples have interesting 
duality properties as well. Using by now familiar arguments 
\cite{fluxbranes}, the last examples are related to orbifolds of the 
conifold considered in \cite{conifold}, with non-trivial embedding of the 
orbifold twist in the RR 1-form gauge degree of freedom.

\section{Freely acting isometries}
\label{freely}

A possibility not explored in the literature is the construction of 
fluxbranes by using freely acting $U(1)$ isometries. The previous examples 
led to fluxbranes with core located, so to speak, at the fixed points of 
the associated isometry. Hence one may incorrectly conclude that freely 
acting isometries would not generate fluxbranes. In fact, the construction 
in Section 2 did not assume the existence of fixed points, and led to 
non-trivial flux configurations in the lower dimensional theory, and no 
obstruction to have well-defined cores for the corresponding flux lumps.

Let us discuss an example of freely acting isometry, namely the Hopf 
isometry for $\IS^3$ with the round metric. Of course string or M-theory 
compactifications on $\IR^{d}\times \IS^3$, for $d=7,8$ respectively, do 
not satisfy the equations of motion. However, the example provides a good 
toy model, which can be embedded in more complicated (and 
interesting) situations, like for instance when the $\IS^3$ is a cycle 
within say a Calabi-Yau space, or for Anti de Sitter compactifications, 
see below.

The round metric for $\IS^3$ can be written
\beqa
ds_3^{\,2} = \sigma_1 \otimes \sigma_1 + \sigma_2 \otimes \sigma_2 + 
\sigma_3 \otimes \sigma_3
\eeqa
where $\sigma_i$ are the $SU(2)$ left-invariant one-forms defined in the 
first column in (\ref{sigmas}). This $\IS^3$ is a (Hopf) fiber bundle 
over $\IS^2$ with fiber $\IS^1$ of constant length. The base is 
parametrized by $\theta, \phi$, while the fiber is parametrized by 
$\psi$, which is an isometric direction with no fixed points.

Introducing an additional flat dimension, parametrized by $x$, the 
metric in the form (\ref{metric}) is
\beqa
ds^2 = dx^2+\sigma_1 \otimes \sigma_1 + \sigma_2 \otimes \sigma_2 + 
(d\psi + \cos\theta\, d\phi)^{\,2} 
\eeqa
Modding by the action $x\to x+2\pi R$, $\psi\to \psi + 4\pi B$, and KK 
reducing, the resulting 3-dimensional configuration is \footnote{Since we 
are not compactifying from 11 to 10 dimensions, one should modify the 
form of the KK ansatz. We skip this point, or if the reader wishes, we 
indeed look at M-theory on say ${\bf X}_{8}\times \IS^3$ (with the radius 
of the 3-sphere a function of some time coordinate in ${\bf X}_8$, to 
account for the instability of the system).}
\beqa
& e^{4\Phi/3}= \Lambda \quad ; \quad A= \Lambda^{-1} (\chi/R) \, 
(d{\tilde\psi} + \cos\theta\, d\phi) & \nonumber \\
& ds^2 = \Lambda^{1/2} (\sigma_1 \otimes \sigma_1 + \sigma_2 \otimes 
\sigma_2) + \Lambda ^{-1/2} (d{\tilde\psi} + \cos\theta\, d\phi)^2 &
\eeqa
with $\chi=2B/R$, and $\Lambda=1+\chi^2$ is a constant. The configuration 
is a 3-sphere, 
squashed on the $\IS^1$ fiber, and preserving the base $\IS^2$ with round 
metric. Moreover there is constant non-zero field strength on the base 
$\IS^2$. The configuration can be thought of as a fluxbrane wrapped on the 
$\IS^1$ fiber (times the space transverse to $\IS^3$ if included). The 
fact that the fluxbrane is smeared over $\IS^2$ is simply  consequence of 
the constant radius of the $U(1)$ isometry, rather than of the lack of 
fixed points, and is avoidable by starting with a lumpy, rather than 
round, metric in $\IS^3$. 

We conclude by discussing a simple context in which a round 3-sphere 
appears, namely in type IIB string compactification on {\bf AdS}$_3\times 
\IS^3\times {\bf X}_4$, with RR 3-form field strength flux through $\IS^3$ 
and where ${\bf X}_4$ a Ricci-flat compact four-manifold. These 
geometries have arisen in the context 
of the AdS/CFT duality as supergravity duals of the theory on D1- and 
D5-branes transverse (resp. wrapped) on ${\bf X}_4$ \cite{maldacena}. Here 
however we look at the configuration in its own right. Considering the 
case ${\bf X}_4=\IT_4$, we may use this 10-dimensional configuration to 
construct a IIB NS-NS fluxbrane in a 9-dimensional compactification. Since 
it is not possible to add a non-compact flat $x_{10}$ direction, we may 
simply use one of the flat (compact) dimensions within $\IT^4$. Denoting 
its coordinate $x$ with periodicity $2\pi R$, we build our 9-dimensional 
fluxbrane by modding by
\beqa
x\to x + 2\pi R/N \quad ; \quad \psi \to \psi + 2\pi n/N
\eeqa
where $\psi$ denotes the above Hopf coordinate, and $n$ is an integer (so 
that the identification we are imposing is consistent with the 
identification implicit in the original torus. The resulting configuration 
is a 9-dimensional {\bf AdS}$_3 \times {\tilde {\bf S}^3}\times \IT^3$.
model, 
with a squashed 3-sphere with constant NS-NS 2-form flux through the base 
$\IS^2$, and (non-constant) RR 3-form flux through $\IS^3$. This can be 
described as a NS-NS flux 6-brane wrapping {\bf AdS}$_3 \times \IS^1\times 
\IT^3$. 

Fluxes have found many applications in the context of AdS/CFT, where they 
correspond in the field theory language to supersymmetry breaking terms, 
like scalar masses, etc. In fact, the above configuration leads to partial 
breaking of supersymmetry, from 8 to 4 supercharges. It would 
be interesting to understand the role of the above construction in the 
AdS/CFT context. 

\section{Type IIA/0A fluxbrane duality}

In \cite{cosgut} the authors proposed a duality of between IIA/0A 
F7-branes for parameters $B$ and $B+1$ (in our convention), respectively. 
Due to the blowing up of the dilaton and the different $B$ parameters, 
a weakly coupled IIA fluxbranes is related to a strongly coupled 0A 
fluxbrane, and viceversa, hence avoiding contradiction with the 
perturbative inequivalence of both theories. 

One may worry that F7-brane solutions with asymptotically finite 
dilaton and vanishing flux built using our construction can lead to 
precisely this kind of contradiction, due to the asymptotic region where 
the theory is weakly coupled regardless of the $B$ parameter controlling the 
flux at the origin. In this section we discuss this issue and solve the 
potential puzzle.

Unfortunately we are not aware of any manifold satisfying the equations of 
motion and admitting a $U(1)$ isometry suitable to obtain F7-branes, and 
with asymptotically constant orbits length. Hence we carry out the 
discussion in terms of a toy model, but which is sufficient to illustrate 
the main point. Interestingly, we find that IIA/0A fluxbrane duality holds 
as in \cite{cosgut} near the fluxbrane core, and breaks down in the 
asymptotic region, precisely where problems could arise. This can be taken 
as mild additional evidence for the proposed duality.

To construct our F7-brane, we consider ${\bf X}_{11}={\bf M}_8 \times 
\IR^2\times \IR$, with metric
\beqa
ds_{11}^2 \, = \, d\vec{x}^{\,2} + d\rho^{\,2} + g(\rho) \, d\varphi^2 + 
dx_{10}^{\,2}
\eeqa
where $g(\rho)$ is a function that interpolates between $\rho^2$ at small 
distance and a constant $R^2$ at large distances, for instance 
$g(\rho)=\rho^2/(1+\rho^2/R^2)$. The ansatz has the drawback that it does 
not satisfy the equations of motion. 
However, one may include time dependence to render it consistent. The 
main point in the following, though, would go through to such situation, 
so we avoid cluttering the discussion.

The $(\rho,\varphi)$ space looks like a (half) cigar, with space 
becoming flat near the tip ($\rho\ll R$), and $\IR\times \IS^1$ away 
from it ($\rho\gg R$). The coordinate $\varphi$ parametrizes a $U(1)$ 
isometry, which is rotational near the tip and has constant orbits of 
length $2\pi R$ in the asymptotic region.

Let us construct a IIA F7-brane by modding out by the identification 
$x_{10}\to x_{10}+2\pi R_{10}$, $\varphi\to \varphi + 2\pi R B$. Reducing 
along $R_{10}\partial_{10}+ B R \partial_{\varphi}$, we obtain
\beqa
& e^{4\Phi/3} \, =\, \Lambda \,= 1+(B\,R/R_{10})^2 \, g(\rho)
\quad ; \quad A_{\tilde\varphi} \, =\, \Lambda^{-1} 
(B\,R/R_{10}^{\, 2})\, g(\rho) & \nonumber \\
& ds_{10}^2 \, = \, \Lambda^{1/2}\, d\vec{x}^{\,2} + \Lambda^{1/2} d\rho^2 
+ 
\Lambda^{-1/2}\, g(\rho)\, d{\tilde\varphi}^2 &
\eeqa
Near the core, the solution looks like the Melvin F7-brane in Section 2, 
while for large $\rho$ the dilaton stabilizes to a finite value, and the 
RR field strength drops. 

The question is the nature of the solution if we reduce along 
$R_{10}\partial_{10}+ (B+1) R \partial_{\varphi}$. For an observer near 
$\rho=0$, things look like in flat space, hence the IIA F7-brane 
with parameter $(B+1)$ should be equivalent to the 0A F7-brane with 
parameter $B$. This is indeed the case because near $\rho=0$ the isometry 
is rotational and fermions are antiperiodic along its orbit, allowing the 
use of \cite{bg} to derive the fluxbrane duality. 

However, if fermions would also be antiperiodic along the $U(1)$ orbit in 
the asymptotic region, we would get to the conclusion that the IIA and 0A 
configurations are equivalent in a region where the flux is negligible and 
the dilaton remains under control. Happily, such a contradiction is 
avoided because in the asymptotic region the fermions are periodic along 
the $U(1)$ orbit. This follows because when the fermions are transported 
along the asymptotic circle, they pick up an additional phase from the 
holonomy of the spin connection. This is given by the exponential of 
the integrated  scalar curvature in the interior of the loop, which by 
the Gauss-Bonet theorem equals $e^{i\pi}=-1$. Hence in the asymptotic region 
both reductions lead to a IIA configuration with no relation to 0A 
fluxbranes.

\section{Conclusions}

In this paper we have proposed a simple extension of the construction of 
the Melvin fluxtube as a quotient of flat space, to build wrapped 
versions of these branes. In particular we have shown that in some 
instances the background curved space confines the flux and leads to 
solutions where the asymptotic fields strength dies off and the dilaton 
becomes constant.

We believe this is an interesting mechanism, worth further exploration, and 
expect these new solutions to help illuminating the still obscure properties 
of fluxbranes. Hopefully, the solutions presented can also lead to useful 
ansatze in the construction of fluxbranes for RR fields other than the 1-form 
(beyond the obvious ones obtained by T-dualizing our examples), perhaps with 
finite flux in the transverse directions (as opposed to certain flat 
fluxbranes of this kind \cite{saffin1,fluxbranes,dielectric}).

Another interesting direction is the study of the world-volume zero modes 
of these objects. In most of our examples the wrapped fluxbranes preserved 
the same symmetries as the background, and led to no obvious goldstone 
zero modes. In a sense, the branes are trapped by the geometry, and are 
simply dragged by the latter is deformed. This does not disagree with 
the fact that the objects are dynamical in better circumstances. We hope 
that further research leads to a more explicit display of dynamical 
features of fluxbranes.

Finally, it would be interesting to introduce D-brane probes in these 
configurations, and learn about D-brane dynamics in new fluxbranes 
backgrounds. In particular, we expect dielectric D-branes to behave better 
in wrapped fluxbranes with asymptotically constant dilatons than in their 
flat space cousins. In particular, the dying off of the flux at large 
distance prevents the dielectric branes from running off to infinity, 
and may lead to stable minima hopefully within the reach of gravitational 
description (see \cite{dielectric,roberto,brecher} for discussion).

\centerline{\bf Acknowledgements}

I am grateful to Roberto Emparan and Jaume Gomis for useful discussions, 
and to M.~Gonz\'alez for kind support and encouragement.

\end{document}

\bibitem{old1}
F.~Dowker, J.~P.~Gauntlett, D.~A.~Kastor, J.~Traschen, `Pair creation of 
dilaton black holes', Phys. Rev. D49 (1994) 2909, hep-th/9309075.